# Evidence for the Active Phase of Heterogeneous Catalysts through *In Situ* Reaction Product Imaging and Multiscale Modeling

S. Matera[1,2], S. Blomberg[3], M.J. Hoffmann[1], J. Zetterberg[4], J. Gustafson[3], E. Lundgren[3], and K. Reuter[1,5,6*]

[1]Chair for Theoretical Chemistry and Catalysis Research Center, Technische Universität München, Lichtenbergstr. 4, 85747 Garching (Germany)
[2]Institute for Mathematics, Freie Universität Berlin, Arminallee 6, 14195 Berlin (Germany)
[3]Division of Synchrotron Radiation Research, Lund University, S-221 00 Lund (Sweden)
[4]Division of Combustion Physics, Lund University, S-221 00 Lund (Sweden)
[5]SUNCAT Center for Interface Science and Catalysis, SLAC National Accelerator Laboratory, Menlo Park, CA 94025 (USA)
[6]SUNCAT Center for Interface Science and Catalysis, Department of Chemical Engineering, Stanford University, Stanford, CA 94305 (USA)

*E-mail: karsten.reuter@ch.tum.de

We use multi-scale modeling to analyze laser-induced fluorescence (LIF) measurements of the CO oxidation reaction over Pd(100) at near-ambient reaction conditions. Integrating density-functional theory based kinetic Monte Carlo simulations of the active catalyst into fluid-dynamical simulations of the mass transport inside the reactor chamber we calculate the reaction product concentration directly above the catalyst surface. Comparing corresponding data calculated for different surface models against the measured LIF signals we can discriminate the one that predominantly actuates the experimentally measured catalytic activity. For the probed CO oxidation reaction conditions the experimental activity is due to pristine Pd(100), possibly coexisting with other (oxidic) domains on the surface.

## 1 Introduction

At the atomic scale, understanding heterogeneous catalysis equates to understanding the specific properties of the catalyst surface atoms in making and breaking chemical bonds. An essential step in this endeavor is to establish the actual surface structure of the operating catalyst, and identify those surface atoms in it that drive the reactions. Reducing complexity to a tractable level, work on single-crystal model catalysts has brought much progress to this end[1,2]. Specifically, this concerns two important insights: (i) The surface of the catalyst can sensitively adapt to the operating conditions. For instance, oxide films may grow on late transition metal catalysts in response to the surrounding oxygen-rich gas-phase environment in oxidation catalysis[3-6]. (ii) This surface must by no means be phase pure. The challenge is thus not only to characterize the surface structure, but also to identify which of possibly different domains predominantly actuate the catalysis.

*In situ* methods to characterize the state of the surface have recently been impressively advanced and begin to provide atomic-scale information at technologically relevant, (near-)ambient pressure conditions[2,5-11]. Reaction-induced compositional and structural changes of the working catalyst are of particular interest. Corresponding studies therefore often specifically target gas-phase conditions leading to highest catalytic activity. Such measurements are then prone to mass transfer limitations (MTLs) and concomitant significant variations in reactant and product concentrations inside the reactor chamber[9,11-15]. In studies on flat-faced model catalysts, this is in particular the formation of a so-called boundary layer of product molecules above the catalyst surface, if these molecules are faster formed by the on-going reactions than can be transported away with the stream in the reactor[14-15]. Such a boundary layer obviously impedes the relation between nominal operation conditions and atomic-scale surface information. Even worse, it also prevents straightforward measurements of the reaction kinetics to further relate this information to catalytic activity[16]. Such measurements generally deduce the catalytic activity through composition analysis of the gas-phase. Samples for which are traditionally extracted through orifices in the reactor wall or in flow reactors simply at the reactor outlet. In the case of MTL-induced concentration variations in the reactor, such samples do not allow for a quantitative activity determination. Neither do they allow to distinguish the source of activity in case of coexisting phases.

One alternative is the sophisticated placement of minimally invasive sampling capillaries[17,18] to obtain the required local and at best spatially resolved information of the gas-phase composition close to the catalyst surface. Another possibility are non-invasive imaging techniques like laser-induced fluorescence (LIF)[19]. In the past such data has already been used to screen for active catalysts[20] to obtain partial mechanistic insight through the detection of radical reaction intermediates[21-23], or to reveal spatio-temporal gradients in reactivity in porous catalyst materials[24]. Here we show that the local kinetic information provided through LIF data can be analyzed to generate much deeper understanding, namely to provide information on the active phase at the operating catalyst surface. This is made possible through the intimate combination with novel multiscale catalysis modeling that integrates predictive-quality first-principles microkinetic simulations into a macroscale description of the detailed mass transport inside the experimentally employed reactor geometry[25]. These calculations yield gas-phase concentration profiles above the catalyst surface for different possible active phase models. Through comparison with the measured LIF signals this provides evidence for the active phase present in the experiment. We demonstrate this approach with the application to CO oxidation at a Pd(100) model catalyst, where the possible formation and role of an oxidic overlayer at (near-) ambient pressure conditions has been controversially discussed[26,27]. Our analysis clearly shows that even under the probed oxygen-rich reaction conditions the high catalytic activity derives predominantly from active sites offered by the pristine metal surface. A new perspective on previous controversies arises as our analysis also suggests that this active phase may not extend over the entire surface, with the remaining area possibly covered by largely inactive oxidic domains.

## 2 Methodology

The LIF experiments were carried out in a stainless steel reaction cell with a total volume of 240 ml[28]. At the reactor inlet the reactant gases are introduced via individual Bronkhorst mass flow controllers that can vary the gas flow from 2 to 50 ml/min. A pressure controller ensures constant total pressure in the reactor. Forming the laser beam into a laser sheet, the LIF measurements are performed in a 2D (planar LIF) mode. The laser is tuned to a wavelength that matches the energy level transition of a gas-phase species of interest and the emitted fluorescence light is recorded with a 2D detector positioned perpendicular to the laser sheet. This generates an image of the concentration of the interrogated species above the catalyst surface with a repetition rate of 10 Hz and a spatial resolution of approximately 400 µm. For CO oxidation we specifically monitor the $CO_2$ production by probing the ((0000) → (10001)) ro-vibrational transition in the $CO_2$ gas molecule at 2.7 µm with a pulse length of 8 ns. Other molecules like CO, NO or $NH_3$ could equally be accessed by LIF either through ro-vibrational transitions such as for $CO_2$ or through electronic transitions. Through electronic transitions LIF also offers the possibility to probe short-lived intermediates such as CH or OH. There are, however, some molecules that are much harder to detect through LIF such as $CH_3$ (which is predissociative) and methanol (where most of the energy is lost in internal energy transfer).

The crystal was cleaned with sputtering and oxygen treatment in an external UHV chamber. The crystal was exposed to air a short time before it was mounted in the reactor. To reduce the resulting contamination on the surface (especially water and hydrocarbons), the crystal temperature was ramped up and down in a CO and $O_2$ environment before the real experiment was performed. From other studies this procedure is known to remove contaminations, as also reflected by insignificant changes in measured LIF-signals when running the temperature-ramp of Fig. 2 twice.

For the microkinetic modelling we relied on first-principles kinetic Monte Carlo (1p-kMC) simulations.[29] The employed 1p-kMC models for CO oxidation at the pristine Pd(100)[30] and at the ($\sqrt{5}\times\sqrt{5}$)$R$27°-O surface oxide[31] have been detailed before and are summarized again for self-containment in the supplementary information (SI). All rate constants entering these models are computed with transition-state theory and density-functional theory (DFT) using either the PBE[32] or RPBE[33] exchange-correlation (xc) functional. The steady-state intrinsic catalytic activity predicted by these models is mapped for a wide range of temperatures, $O_2$ and CO partial pressures, and is subsequently interpolated using a modified Sheppard algorithm[25]. This continuous representation then serves as the boundary condition representing the single-crystal catalyst in the fluid-dynamical simulations. The latter simulations are performed with the CatalyticFoam package[34] for a detailed mesh model of the experimental reactor chamber as detailed in the SI. During the temperature ramp, the flow pattern is assumed to have equilibrated at each temperature. With a linear LIF signal to $CO_2$ concentration relation, calibration of the calculated data is finally achieved by normalizing to the LIF signal in the saturated MTL plateau obtained for the Pd(100) model at 600 K. Further details are provided in the SI.

## 3 Results

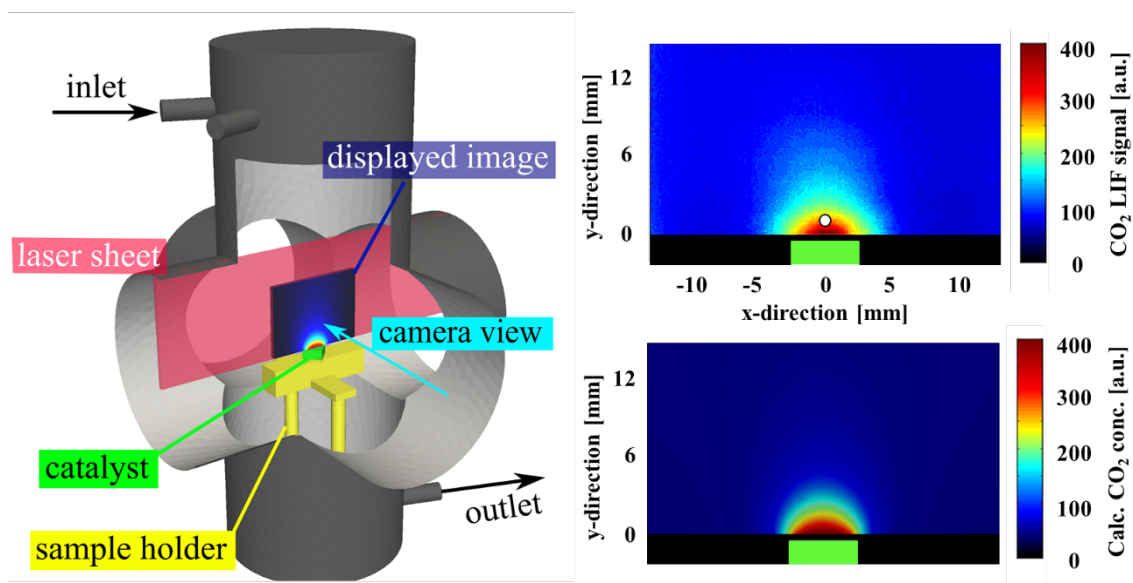

Figure 1: (Left panel) Representation of the reactor employed in the laser-induced fluorescence (LIF) measurements. A 2D laser sheet expands perpendicular to the Pd(100) surface and the LIF signals are measured in the highlighted 15x27.5 mm² rectangle above the catalyst. (Right upper panel) Measured LIF image reflecting the CO2 gas-phase distribution for an oxygen-rich gas feed with a 4:1 $O_2$:CO ratio and a catalyst temperature of 600 K (total pressure: 180 mbar, 50% Ar, inlet mass flow: 72 mln/min). The small white circle in the center of the image indicates the area over which the integrated LIF signal is analyzed in Fig. 2. (Right lower panel) Simulated $CO_2$ concentration profile for the same feed conditions and using Pd(100) in its pristine metal state as model for the active phase (see text).

Figure 1 shows a representation of the reactor employed in the LIF experiments[28]. The reactant gases enter at the reactor inlet at the upper left, flow across the centrally placed, flat-faced Pd(100) model catalyst and exit the reactor at the lower right. The LIF measurements monitor the catalytic activity by probing the $CO_2$ product concentration in a planar sheet perpendicular to the catalyst surface as illustrated in Fig. 1. Previous problems in accessing this fruit-fly molecule of model catalysis with LIF are hereby overcome by an orders of magnitude improved laser power. The right panel in Fig. 1 shows a corresponding LIF image obtained for an oxygen-rich gas feed with a 4:1 ratio of $O_2$:CO, at a near-ambient total pressure of 180 mbar and a catalyst temperature of 600 K. These reaction conditions are motivated by previous *in situ* X-ray photoelectron spectroscopy (XPS) work[9]. For a similarly O-rich gas-phase composition a thin surface oxide film was identified as predominant phase on the surface in that work – at a high, but not further quantifiable catalytic activity[9].

The LIF image confirms a very high catalytic activity also in the present case: High enough to lead to strong MTLs as apparent from the pronounced boundary layer of $CO_2$ product molecules above the surface. In this MTL-controlled regime no reliable kinetic information could indeed be obtained by standard gas-phase compositional analysis at the reactor outlet or close to the reactor walls. The LIF reaction product imaging, on the other hand, allows to probe the catalytic activity directly at the catalyst surface. Despite the MTLs we can thus use it to identify the surface phase that predominantly actuates this activity. For this we employ our multiscale modeling approach, in which we integrate the DFT-based 1p-kMC microkinetic model of the active phase into macroscale flow simulations in the experimental reactor geometry[25]. This allows to predict the detailed $CO_2$ concentration profile above the catalyst surface for different possible active phase models. With a linear LIF signal to $CO_2$ concentration relation, agreement of the calculated signature for an active phase model with the experimental data provides then indirect evidence that the corresponding phase is predominantly responsible for the measured catalytic activity. Specifically and as shown in Fig. 1, we obtain a boundary layer of equal shape and extension when we use as the active phase actuating the catalytic activity the established 1p-kMC model for CO oxidation at Pd(100) in its pristine metal state[30]. In making this statement we hereby disregard the lowest LIF signal noise and focus on the core of the boundary layer close to the surface.

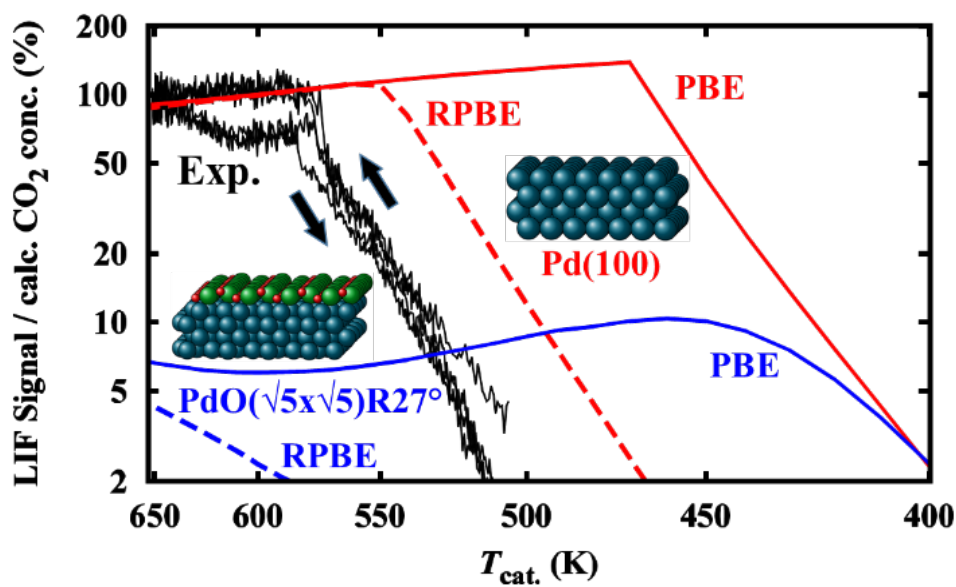

Figure 2: Measured LIF signal integrated over the small area highlighted in Fig. 1 (solid black line). Shown is a ramp of the catalyst temperature from 500 K to 650 K and back (as indicated by the arrows), for otherwise identical feed gas conditions as in Fig. 1 (4:1 $O_2$:CO ratio, total pressure: 180 mbar, 50% Ar, inlet mass flow: 72 mln/min). Additionally shown is the corresponding calculated $CO_2$ concentration variation as predicted for the ($\sqrt{5}\times\sqrt{5}$)$R27°$-O surface oxide (blue lines) and for the pristine metal state of Pd(100) (red lines). To assess the uncertainties arising from the approximate DFT energetics, data obtained with the PBE[29] (solid lines) and RPBE[30] (dashed lines) xc functional are shown. Apart from a shift in onset temperature, only the signature computed for the pristine metal state can be reconciled with the experimental data, see text. Insets show representations of the surface oxide and Pd(100) structures (Pd: dark green spheres, O: red spheres, Pd atoms inside the surface oxide layer are in light green).

While generally viable, one has to recognize though that for highest catalytic activity the extension of the then fully developed product boundary layer is more sensitive to the reactor geometry and its fluid flow pattern than to the actual catalyst[35]. Any catalyst state with a high enough intrinsic catalytic turnover would yield a similar $CO_2$ profile. Rather than on the detailed spatial distribution for one set of reaction conditions we therefore focus on a range of different reaction conditions. Each time we only monitor the LIF signal viz. $CO_2$ concentration integrated over a small area directly above the catalyst surface as illustrated in Fig. 1. Interested primarily in less active conditions where the boundary layer is not yet saturated, we specifically perform ramps of the catalyst temperature for otherwise constant inlet gas-phase composition.

Figure 2 shows the corresponding LIF signal during a linear temperature ramp that goes from 500 K to 650 K and back for the same reactant feed as in Fig. 1. No influence on the obtained curve was observed when doubling the heating rate from 0.15 K/s to 0.30 K/s. A strong increase in the signal with temperature reflects the increasing catalytic activity, until at around 575 K saturation is reached. Increasing and decreasing part of the temperature ramp curiously reflect a small hysteresis, which our fluid-dynamical simulations rationalize in terms of a delayed warm-up of the reactor walls. The higher overall temperature in the decreasing part of the ramp then slightly reduces the gas-phase density and therewith the LIF signal.

Using again the 1p-kMC model for Pd(100) in its pristine metal state[30] as the active phase in our multiscale modeling, we obtain an overall reaction product signature above the catalyst that nicely matches the experimental one. As shown in Fig. 2 we reproduce the linear increase in activity until a saturation plateau is reached, with an apparent activation barrier (i.e. slope of the linear regime) that agrees very well with experiment. The entire signature is shifted by about 100 K to lower temperatures though, which motivates to assess alternative candidates for the active phase in the simulations. Most obvious would hereby be the ($\sqrt{5}$x$\sqrt{5}$)$R$27°-O surface oxide reconstruction[36] that had been characterized as predominant surface phase in the preceding *in situ* XPS work for an equally oxygen-rich gas phase composition[9]. However, when using the established 1p-kMC model for this surface oxide[31] in the simulations, the calculated signature disagrees qualitatively with the measured LIF data, cf. Fig. 2. The lower and only weakly temperature-dependent activity can hereby readily be rationalized by the much weaker CO binding on the surface oxide compared to Pd(100)[30,31]. At the higher temperatures of the ramp, we correspondingly find the catalytic activity of the surface oxide limited by the largely depleted on-surface CO population atop of the oxide film.

## 4 Discussion

For a given active phase model the largest uncertainty in the simulations arises from the approximate DFT energetics underlying the 1p-kMC model. A detailed sensitivity analysis of the Pd(100) 1p-kMC model by systematically varying individual rate constants away from their DFT values shows that the CO oxidation reaction constitutes the rate-determining step at the probed reaction conditions, whereas for the ($\sqrt{5}$x$\sqrt{5}$)$R$27°-O surface oxide model adsorbate binding energies are most critical. We expect the employed DFT-PBE[32] exchange-correlation functional to rather be on the over-binding side. Correspondingly, we rerun the simulations describing these critical processes[31,36] at the level of the much weaker binding DFT-RPBE[33] functional. For the surface oxide this weakened bonding further aggravates the problem of stabilizing CO at the surface and we now obtain an essentially negligible catalytic activity over the conditions of the temperature ramp, cf. Fig. 2. For the Pd(100) we instead obtain a signature as before, but shifted by ~50 K to higher temperatures and therewith to closer agreement with the experimental signature. Even in light of the DFT uncertainty the computed signature of the ($\sqrt{5}$x$\sqrt{5}$)$R$27°-O surface oxide model can thus not be reconciled with the experimental data. In contrast, the Pd(100) model does yield a signature that is highly compatible with the experimental data, where the small differences in the onset temperature could be attributed to the approximate DFT energetics. Despite the O-rich gas-phase composition of 4:1 in $O_2$:CO pressure we are therefore forced to conclude that the phase that dominates the catalytic activity under the probed reaction conditions is Pd(100) in its pristine metal state.

Quantitative agreement with the experimental data would be reached when increasing the CO oxidation barrier in the Pd(100) 1p-kMC model to a value that is even higher by 0.1 eV than the DFT-RPBE value. A corresponding inaccuracy is generally well within the range that has to be expected for prevalent generalized-gradient functionals. On the other hand, the RPBE functional is already on the weaker binding side. Alternative to conjecturing a further increased barrier to rationalize the remaining disagreement with experiment, it is thus appealing to recall that the probed reaction conditions fall close to the stability boundary between the pristine metal and the oxidized state[4,31]. Operating close to this boundary, a coexistence of Pd(100) and ($\sqrt{5}$x$\sqrt{5}$)$R$27°-O surface oxide domains on the catalyst surface is then well conceivable. Disregarding any special catalytic activity at domain boundaries, a coexistence of mesoscopic domains can be accounted for in the flow simulations by linearly mixing the catalytic activities predicted from the Pd(100) and surface oxide 1p-kMC model. At the DFT-PBE level such a mixing does not lead to any improvement with respect to the measured signature, as both models predict a similar onset of catalytic activity at temperatures well below experiment, cf. Fig. 2. In contrast, when employing the DFT-RPBE energetics the surface oxide exhibits a negligible activity. The signature obtained when assuming that the active metal phase covers only a fraction $x < 100\%$ of the entire surface therefore simply corresponds to a scaled-down version of the pure Pd(100) metal signature in Fig. 2: The linear increase starts at a higher onset temperature but with unchanged slope, such that the saturation plateau is also reached at a correspondingly higher temperature. For a fraction $x \approx 25\%$ we can therefore also reach quantitative agreement with the experimental LIF data.

## 5 Summary and Conclusions

In summary, we have used LIF as a local, non-invasive and *in situ* probe of the reaction kinetics. We have analyzed the obtained data with a multiscale modeling approach that integrates first-principles microkinetic models of different possible active phases into fluid-dynamical simulations. This provides indirect evidence which of these phases is predominantly responsible for the measured catalytic activity. In the present case of near-ambient CO oxidation at Pd(100), only the calculated signature for the pristine metal state is found compatible with the measured data – despite the employed O-rich reaction conditions. Remaining differences in the calculated and measured onset temperature for the catalytic activity could arise from inaccuracies in the DFT energetics underlying the simulations. Alternatively, they could arise from a heterogeneous surface in which the active metal domains form only a minority phase. Without further characterization of the surface, we cannot distinguish either of these two possible rationalizations on the basis of the present multiscale modeling alone. What is clear though is that the dominant activity comes from the pristine metal state, regardless whether the surface is homogeneously covered by a metal termination or whether it exhibits a phase mixture with metal domains.

A rationalization in terms of a phase mixture could hereby potentially resolve many past controversies in the *in situ* characterization field, where in the absence of clear-cut kinetic information no distinction could be made between phases that are predominantly present at the surface and phases that are predominantly responsible for the catalytic activity. In this respect, the novel LIF analysis presented here will form a most valuable addition. LIF can readily be combined with a wide range of *in situ* characterization techniques. This combination will then provide simultaneous information on structure, composition and reaction kinetics, and will thus represent a major step towards the ultimate goal of unambiguously identifying the active phases of working catalysts.

## Acknowledgements

S.M., M.J.H. and K.R. gratefully acknowledge funding from the German Research Council (DFG). S.B., J.Z., J.G. and E.L. gratefully acknowledge funding from the Swedish Research Council (VR) and the Swedish Foundation for Strategic Research (SSF). J.Z. thanks the Knut and Alice Wallenberg Foundation.

Supporting information are provided with details on the computational approach. This information is available free of charge via the Internet at http://pubs.acs.org/

**Graphical TOC entry**

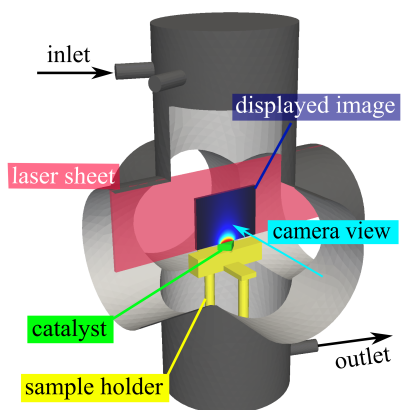